\documentclass[%
 reprint,
 amsmath,amssymb,
 aps,
 prl,%longbibliography,%onecolumn
]{revtex4-1}

\usepackage{graphicx}
\usepackage{dcolumn}
\usepackage{bm}
\usepackage{color}
\begin{document}

%\title{  Nonscattering  Large Purcell Enhancement Induced by Topological Photonic Structures}
%\title{  Nonscattering  Large Purcell Enhancement caused by Topological photonic Structures}
%\title{  Nonscattering  Large Purcell Enhancement resulted fromTopological photonic Structures}
%\title{ Topologically Enabled Ultra-high-Q Guided Resonances Robust to Out-of-plane Scattering }
\title{ Topologically Enabled Ultralarge Purcell Enhancement Robust to Photon Scattering }

\author{Zhiyuan Qian$^1$}
\author{Zhichao Li$^1$}
\author{He Hao$^1$}
\author{Lingxiao Shan$^1$}
\author{Qihuang Gong$^{1,2,3,4}$}
%\author{C. T. Chan$^5$}
\author{Ying Gu$^{1,2,3,4}$}
 \email{ygu@pku.edu.cn}

\affiliation{$^1$State Key Laboratory for Mesoscopic Physics, Department of Physics, Peking University, Beijing 100871, China\\
$^2$Nano-optoelectronics Frontier Center of the Ministry of Education $\&$  Collaborative Innovation Center of Quantum Matter, Peking University, Beijing 100871, China\\
$^3$Collaborative Innovation Center of Extreme Optics, Shanxi University, Taiyuan, Shanxi 030006, China\\
$^4$Beijing Academy of Quantum Information Sciences, Beijing 100193, China\\
%$^5$Department of Physics, Hong Kong University of Science $\&$ Technolology, Clear Water Bay, Hong Kong, China
}

\date{\today}
\begin{abstract}

Micro/nanoscale single photon source is a building block of on-chip quantum information devices.
Owing to possessing ultrasmall optical mode volume, plasmon structures can provide large Purcell enhancement, however scattering and absorption are two barriers to prevent them from being used in practice.
To overcome these barriers, we propose the topological photonic structure containing resonant plasmon nanoantenna, where nanoantenna provides large Purcell enhancement while topological photonic crystal guides all scattering light into its edge state.
Through the optical mode design, the rate of single photons emitted into the edge state reaches more than $10^4\gamma_0$ simultaneously accompanied with an obvious reduction of absorption.
This kind of nonscattering large Purcell enhancement will provide new sight for on-chip quantum light sources such as a single photon source and nanolaser.
 
\begin{description}
\item[Keywords]  Topological photonic structrue; Plasmon nanoantenna;  Purcell enhancement
%\item[PACS numbers]
\end{description}

\end{abstract}

%\pacs{Valid PACS appear here}

\maketitle

%%%% Section I
\section{1.  Introduction}
Micro/nanoscale single photon source is an indispensable building block of on-chip quantum information processing \cite{single-photon-1, single-photon-2}.
Utilizing local field enhancement or high density of optical modes of photonic structures to improve the spontaneous emission of single emitter is one of key principles of realizing a single photon emission.
Typical micro/nanostructures include whispering guided resonantors \cite{WGM1}, photonic crystals (PCs)\cite{CQED_PC1,CQED_PC2, CQED_PC3}, and plasmon nanostructures \cite{CQED_MNP4,CQED_MNP5,CQED_MNP6}.
Though achieving large Purcell enhancement, the  volume of  whispering guided resonantor itself is at the microscale  \cite{WGM1}, which to some extent prevents compact integration of on-chip.
PC cavities can effectively  increase the photon emission, but the emission rate is generally low \cite{CQED_PC1,CQED_PC2, CQED_PC3}.
Owing to possessing ultrasmall optical mode volume, plasmon nanostructures can provide large Purcell enhancement \cite{CQED_MNP4,CQED_MNP5,CQED_MNP6}.
However, their scattering and absorption are two barriers when guiding these  single photons into other devices.
To solve the problem of scattering,  the gap surface plasmon structures are proposed by combining the advantages of effectively collecting scattering light by a nanowire or nanofilm and inducing large Purcell enhancement by plasmon nanoparticles \cite{Gu-1,Gu-2,Gu-3}.  
However, their collecting efficiency is not very high and it is almost impossible to reach 100$\%$ collecting of photons. Thus, the final part guided into other on-chip devices is very low and those stray light will severely affect the performance of neighbouring devices, which prevent them from being a good candidate to realize a high quality single photon source.

By introducing quantum Hall effect into optics, topological photonics becomes an important branch in micro/nano photonics \cite{1,2,4}.
Topological states are some specific optical modes existing between the optical bands, characterized as topological invariants  in the reciprocal space \cite{chan-1,3,5}, such as Berry curvature, Berry phase, and Chern number.
So those micro/nano structures with energy bands and gaps, such as photonic crystals \cite{6}, coupled-resonator arrays \cite{7}, and synthetic dimensions spatial-modal lattices \cite{8}, are good candidates to realize the topological properties. 
Topological states, generally referring to edge states or interface states, are characterized as nonscattering propagation of photons and immunity to a wide class of impurities and defects, i.e.,  topological protection  \cite{3,5}.
These features allows to fabricate various micro/nano photonic devices, including topological lasers  \cite{9,10,11}, nonscattering sharp bent waveguides  \cite{12,13}, and topological quantum light \cite{14}.
Recently, topological protection is utilized in on-chip quantum information processing, such as robust transport of
entangled photons \cite{15}, protection of biphoton states \cite{16}, and topological phase transition in single-photon dynamics  \cite{17}.
However, using topological protection into the Purcell enhancement has not been reported yet.
Though overcoming the scattering problem in the propagation process,  if only depending on edge states to improve Purcell enhancement, its emission rate  can not be very high.
 
To the end of obtaining nonscattering Large Purcell enhancement, we propose a specific topological photonic structure, i.e., 1D topological PC containing a resonant nanoantenna [Fig. 1(a)].
By embedding an antenna into the topological PC, strong local field near antenna leads to a large Purcell enhancement while nonscattering edge state can make all scattering photons propagate along some specific direction.
In condition of topological protection, through the frequency matching between resonant antenna and edge state, total Purcell factor can reach more than $2\times10^4\gamma_0$ ($\gamma_0$ is the spontaneous emission rate in vacuum), among which the propagating part along the edge state channel is more than $10^4\gamma_0$.  
Interestingly,  there is an obvious absorption reduction in total Purcell enhancement due to the deformation of nearfield of nanoantenna caused by the edge state. 
Moreover, the Purcell factors are very sensitive to the nanoantenna size, relative position of nanoantenna and quantum emitter in topological PC. 
Through propagating those single photons  along the edge state,  scattering problem  is solved, which will be directly  used to on-chip quantum devices. 
This kind of nonscattering ultralarge Purcell enhancement will provide new sight for on-chip quantum light sources such as a single photon source and nanolaser.

%%%% Section II
\section{ 2. Optical modes of topological photonic structures}

We choose 1D topological photonic structure composed of two semi-infinite PCs  with layers A and B [Fig. 1(a)]. 
This kind of topological PC, characterized as  Zak phase, was proposed by Chan et al. \cite{chan-1}.
And then the concept of Zak phase was extended to various 2D photonic structures \cite{chan-2,chan-3,chan-4,chan-5} and applicated in topological light-trapping \cite{NC-trapping} and nanolaser \cite{CP}.
However, nonscattering propagation in edge state has not been brought into the Purcell enhancement.
The edge state in a dielectric PC is a bulk mode, so its Purcell factor can not be very large, generally dozens of $\gamma_0$ \cite{supple}. While resonant metallic nanopartice with strong local field can take great Purcell enhancement  \cite{CQED_MNP4,CQED_MNP5,CQED_MNP6}. 
However, due to the loss of absorption and scattering, the single photons dispersing around the nanoparticle is difficult to be utilized.
Therefore, by combining the advantages of the topological PC and plasmon nanoparticle, we design the topological photonic structure containing 1D PC and metallic nanoantenna [Fig. 1(a)].
With this structure, we obtained large Purcell factors, among which all scattering part of single photons is guided into the edge state.
Owing to the  small size of nanoantenna,  in the following computation, the topological protection of whole structure is conserved \cite{supple}.

\begin{figure}
  \centering
  % Requires \usepackage{graphicx}
  \includegraphics[width=0.5\textwidth]{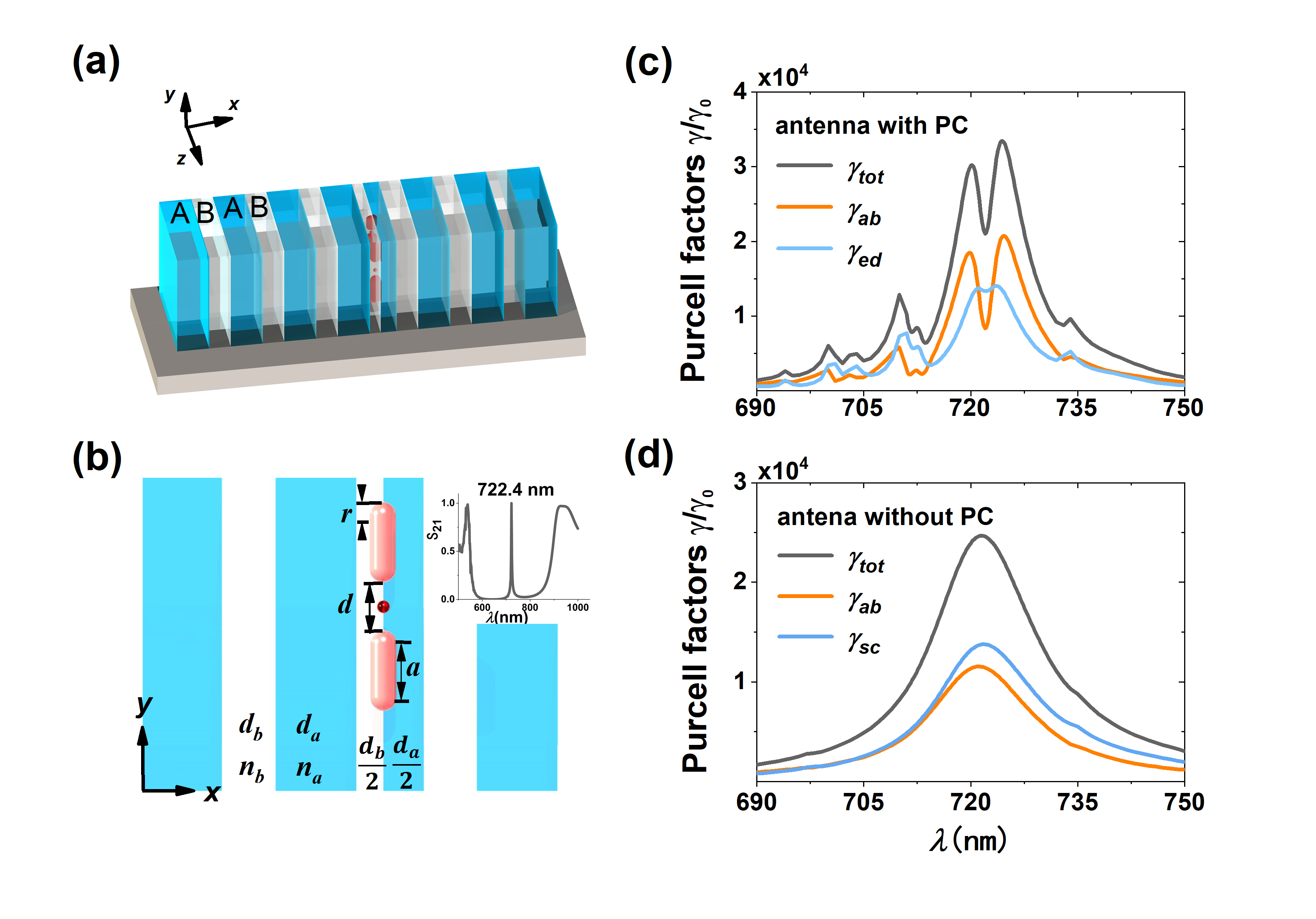}\\
  \caption{ Nonscattering Purcell enhancement in topological photonic structure.
(a) Schematic diagram and (b) its section view of topological photonic nanostructure composed of a 1D topological PC, a silver nanoantenna,  and a quantum emitter. 
Purcell factors at the gap center of nanoantenna (c) with and (d) without the designed topological photonic structure as a function of wavelength $\lambda$. 
It is found that the nonscattering part $\gamma_{ed}/\gamma_0$ of Purcell factor in topological photonic structure is more than $10^4$. While the nanoantenna is in the dielectric, the nonabsorption part $\gamma_{sc}/\gamma_0$ is also very large, but all of them scatters into the free space.  }\label{fig1}
\end{figure}

If now putting a quantum emitter into the near field region of nanoparticle, total Purcell enhancement can be divided into three parts, i.e., $\gamma_{tot}=\gamma_{ab}+\gamma_{ed}+\gamma_{sc}$, where 
$\gamma_{ab}$ is the absorption part, $\gamma_{ed}$ the decay rate into the edge state, and $\gamma_{sc}$ radiative part into free space. 
It is different from that the metallic nanoparticle is embedded in the dielectric, where $\gamma_{tot}=\gamma_{ab}+\gamma_{sc}$, among which  $\gamma_{sc}$ is the scattering part into the free space and can not be effectively collected \cite{CQED_MNP4,CQED_MNP5,CQED_MNP6}. 
In topological PC, if the metallic nanoparticle is small enough, $\gamma_{sc}$ can be neglected due to topological protection from impurities.  
For our designed topological photonic structure, the part of $\gamma_{sc}$ is totally suppressed while the part of $\gamma_{ed}$ collects all the scattering photons \cite{supple}, i.e., $\gamma_{tot}=\gamma_{ab}+\gamma_{ed}$ is validated. Thus single photons from the part $\gamma_{ed}$ can be used in the on-chip photonic devices with nonscattering. 

As shown in Fig. 1(b), the topological 1D PC consists of two semi-infinite PCs of two layers A and B with the thickness of $d_a$=120 nm and $d_b$=100 nm and the refractive index of $n_a$=2 and $n_b$=1.
With these parameters,  the edge state appears at the wavelength of $\lambda=$722.4 nm shown as the inset of (b). 
The nanoantenna with the  resonant wavelength of $\lambda=$722.4 nm is horizontally placed into the interface of two semi-infinite PCs. 
The spectral linewidth of edge state and surface plasmon of nanoantenna are 2.7 nm and 18.8 nm respectively.
A quantum emitter is set at the nanoscale gap of antenna.
To compute the Purcell factors of above topological structure, 3D finite-element simulations were performed using COMSOL Multiphysics software, through which we have  simulated optical modes, Purcell enhancement, photon collection, and photon-emitter coupling strength for various photonic structures  \cite{Gu-NN, Gu-1,Gu-2, Gu-3,Gu-4,Gu-5}. 
To simulate the infinite 1D PC, periodic boundary condition is applied for the vertically directional boundary of propagation. In the direction of photonic propagation, 5 periods for both sides of semi-infinite photonic crystal are enough to perform infinite-like behavior. The emitter is represented by a polarized dipole point source. 
Computation details of $\gamma_{tot}$, $\gamma_{ab}$,  $\gamma_{sc}$, and  $\gamma_{ed}$ are shown in Ref. \cite{supple}.
In the supplementary materials, we have proved no scattering part through the four boundary of propagation because of the topological protection as well as the correctness and validity of above module.

%%%% Section III
\section{ 3. Results and discussions}
\subsection{ 3.1 Nonscattering ultralarge Purcell enhancement at edge state }

Nonscattering Purcell enhancement of topological PC containing the Ag nanoantenna is investigated.
 As shown in Fig. 1(c), total Purcell factor $\gamma_{tot}/\gamma_{0}$ can reach more than $2\times10^4$, among which the part $\gamma_{ed}/\gamma_{0}$ guided into the edge state is $10^4$. Here, the parameters of silver nanoantenna are $r=$7 nm, $a=$24.45 nm, and $d=$10 nm.
 Owing to the topological protection, there is almost not any photon scattering.
 At the edge state of $\lambda$=722.4 nm, which is also corresponding to the resonant wavelength of Ag nanoantenna, there is a dip in the spectra of $\gamma_{tot}/\gamma_{0}$ and $\gamma_{ab}/\gamma_{0}$, and  
$\gamma_{tot}/\gamma_{0}$ is in its minimum while the $\gamma_{ed}/\gamma_{0}$ reaches its maximum.
This ultralarge nonscattering enhancement is superior to that of gap surface plasmon structures, where the Purcell factor is also very large, but the guided part is relatively small and the stray light exists \cite{Gu-1,Gu-2,Gu-3}. 
The linewidth of $\gamma_{ed}/\gamma_{0}$ is greatly less than that of $\gamma_{tot}/\gamma_{0}$ due to the decoupling bewteen the 1D PC and Ag nanorod.
It is also found that, at the edge state, the ratio of $\gamma_{ab}/\gamma_{tot}$ (37.8$\%$)  in topological structure is less than that of $\gamma_{ab}/\gamma_{tot}$ (45.6$\%$) in dielectric [Figs. 1(c, d)], which means an obvious absorption reduction due to the existence of edge state.

\begin{figure}
  \centering
  % Requires \usepackage{graphicx}
  \includegraphics[width=0.5\textwidth]{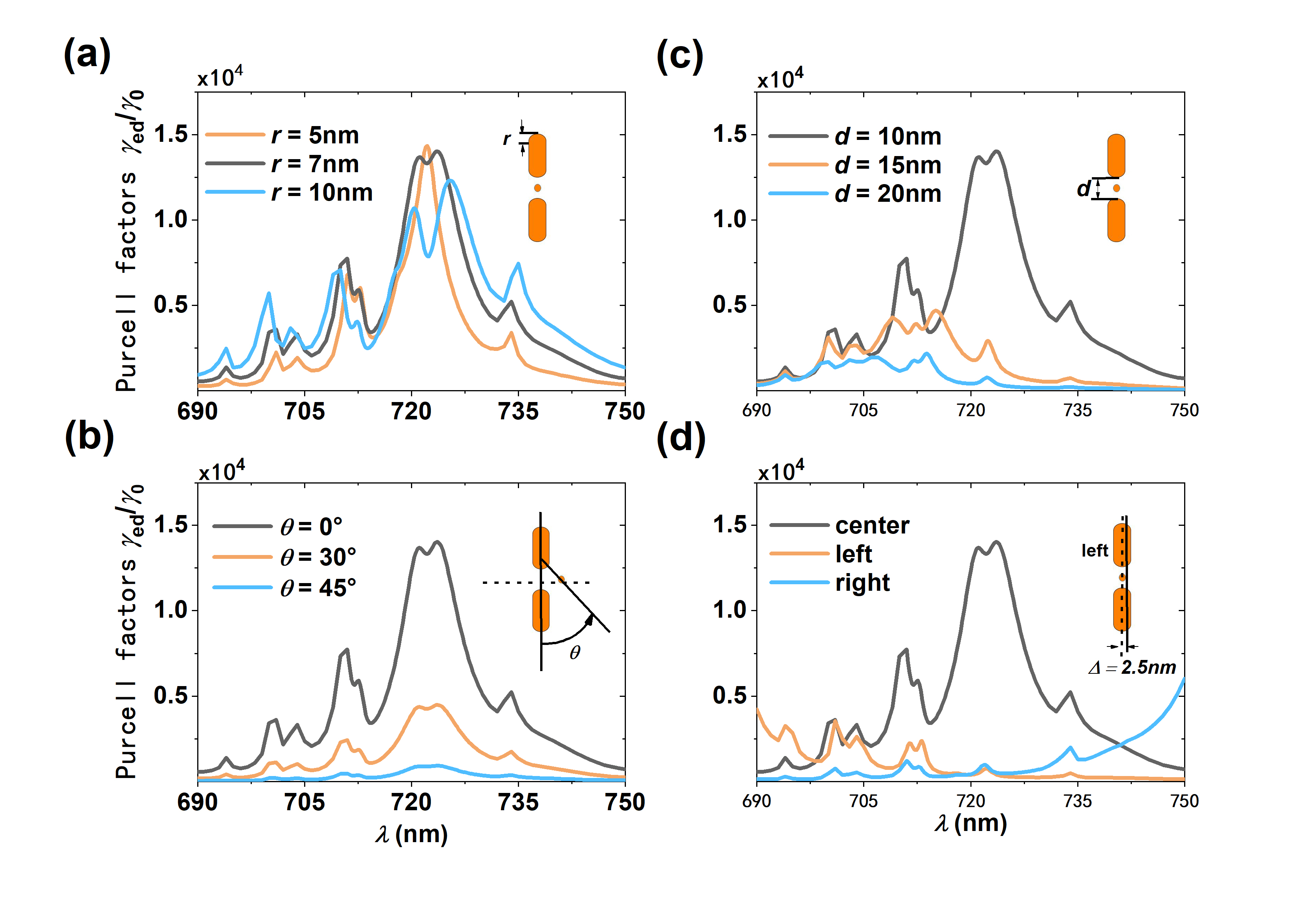}\\
  \caption{ Nonscattering Purcell factors  $\gamma_{ed}/\gamma_{0}$ for various parameters of nanoantenna embedded in topological PC. 
 (a) Purcell factors $\gamma_{ed}/\gamma_{0}$ with different radius $r$ of nanoantenna as a function of $\lambda$.
  Here the distance $d$ between the nanorods is fixed at $d=$10 nm. Other parameters are $r=5$ with $a=18.9$ nm,  $r=7$ nm with $a=24.5$ nm, and $r=10$ nm with $a=30.5$ nm, separately. 
 % (b) Purcell factors $\gamma_{ed}/\gamma_{0}$ with different distance $d$ as a function of $\lambda$.
% To match the resonant condition between the nanoantenna and edge state,  the radius and length of nanorod are $d=10$ with $r=?$ and $a=??$ nm,  $d=15$ nm with $r=?$ and $a=??$ nm, and $d=15$ nm with $r=?$ and $a=??$ nm, separately.
 % Nonscattering part $\gamma_{ed}/\gamma_{0}$  becomes larger with the smaller size or closer distance between nanorods due to more localized near field. 
Purcell factors $\gamma_{ed}/\gamma_{0}$  (b)  with varying the position of quantum emitter, (c) with different distance $d$,  and (d) with moving the position of nanoantenna as a function of $\lambda$. Other parameters are the same as those in Fig. 1(c).
}\label{fig2}
\end{figure}

Then, we focus our attention on the nonscattering Purcell factors $\gamma_{ed}/\gamma_{0}$ for various parameters of nanoantenna embedded in topological PC.
While keeping the resonance matching between the edge state and resonant antenna, let us explore the effect of varying nanorod radius $r$ and emitter position on $\gamma_{ed}/\gamma_{0}$.
When the radius $r$ is smaller, both $\gamma_{tot}$ and $\gamma_{ed}$ become larger due to  more localized electric field [Fig. (2a)].
We also changed the relative angle $\theta$ between the emitter and nanoantenna [Fig. (2b)].
It is found that away from near field region, such as $\theta=45^\circ$, $\gamma_{ed}$ decreases abruptly.
This is similar to the case of resonant metallic nanorod embedded in the dielectric medium,  where the position of the emitter is very sensitive to the scattering part $\gamma_{sc}/\gamma_{0}$ \cite{CQED_MNP4,CQED_MNP5,CQED_MNP6}.
In the near field region, the ratios of $\gamma_{ed}/\gamma_{tot}$ in topological structure  and $\gamma_{sc}/\gamma_{tot}$ in dielectric medium keep almost as a constant respectively. 
But away from the near field region, these ratios as well as $\gamma_{tot}$ descend abruptly. More  computation details are shown in the Ref. \cite{supple}.
Therefore, by changing the structure parameters of nanoantenna or the position of emitter, the magnitude of nonscattering part 
$\gamma_{ed}/\gamma_{0}$ as well as $\gamma_{tot}/\gamma_{0}$ can be well modulated.

Owing to narrow linewidth of the edge state, Purcell enhancement is very sensitive to resonance matching between the nanorod and edge state.
When the gap distance $d$ of nanoantenna is changed from 10 nm to 15 nm, $\gamma_{ed}/\gamma_{0}$ decreases greatly due to  off resonance condition [Fig. 2(c)].
If we slightly move the position of nanoantenna away from the the center of two PC layers, there is a great decrease in $\gamma_{ed}$ due to the red and blue shift of resonance wavelength [Fig. 2(d)].
Thus, Purcell enhancement  in this topological photonic structure is in a large extent determined by the resonance matching.

With the resonance matching condition, we also study the Purcell enhancement in this topological PC containing other kinds of nanoparticles, such as a single  nanorod lying parallel to the interface, a single  nanorod being vertical to the interface, and the  antenna being vertical to the interface \cite{supple}. 
It turns out that as long as the topological protection is conserved, the equation $\gamma_{total}$=$\gamma_{ed}$+$\gamma_{ab}$ is satisfied and large nonscattering Purcell enhancement is maintained. 
Overall, structures of antenna have a larger Purcell enhancement than those of single nanorod, and structures that lies parallel to the interface have larger ratio of $\gamma_{ed}$/$\gamma_{total}$. 
It means that the edge state can greatly change the local field of nanoparticle and then changed local field further influences the  Purcell factors $\gamma_{tot}$, $\gamma_{ed}$ and $\gamma_{ab}$.
These results are helpful to  experimental realization of this protocol.

\subsection{ 3.2 Absorption reduction induced by edge state}

Let us look back to the dip (corresponding to edge state) in the spectra of Purcell factors $\gamma_{tot}/\gamma_{0}$ and $\gamma_{ab}/\gamma_{0}$ in Fig. 1(c).
As shown in Figs. 3(a, b), both the electric field and energy flux for the nanoantenna in the topological structure are drawn away by the edge state comparing with the case of the nanoantenna in the dielectric [Figs. 3(c, d)].
In topological structures, the near field is not localized as that without PC, thus a little decreased $\gamma_{tot}$  is obtained.
According to the electromagnetic boundary condition, in this case the electric field of nanoantenna becomes weaker than that  in dielectric medium, leading to the reduction of absorption.
Simultaneously, the energy flux in topological structure is flowing along the PC directionally [Figs. 3(b)], while the energy flux  in dielectric medium scatters into all directions  [Figs. 3(d)].

\begin{figure}
  \centering
  % Requires \usepackage{graphicx}
  \includegraphics[width=0.5\textwidth]{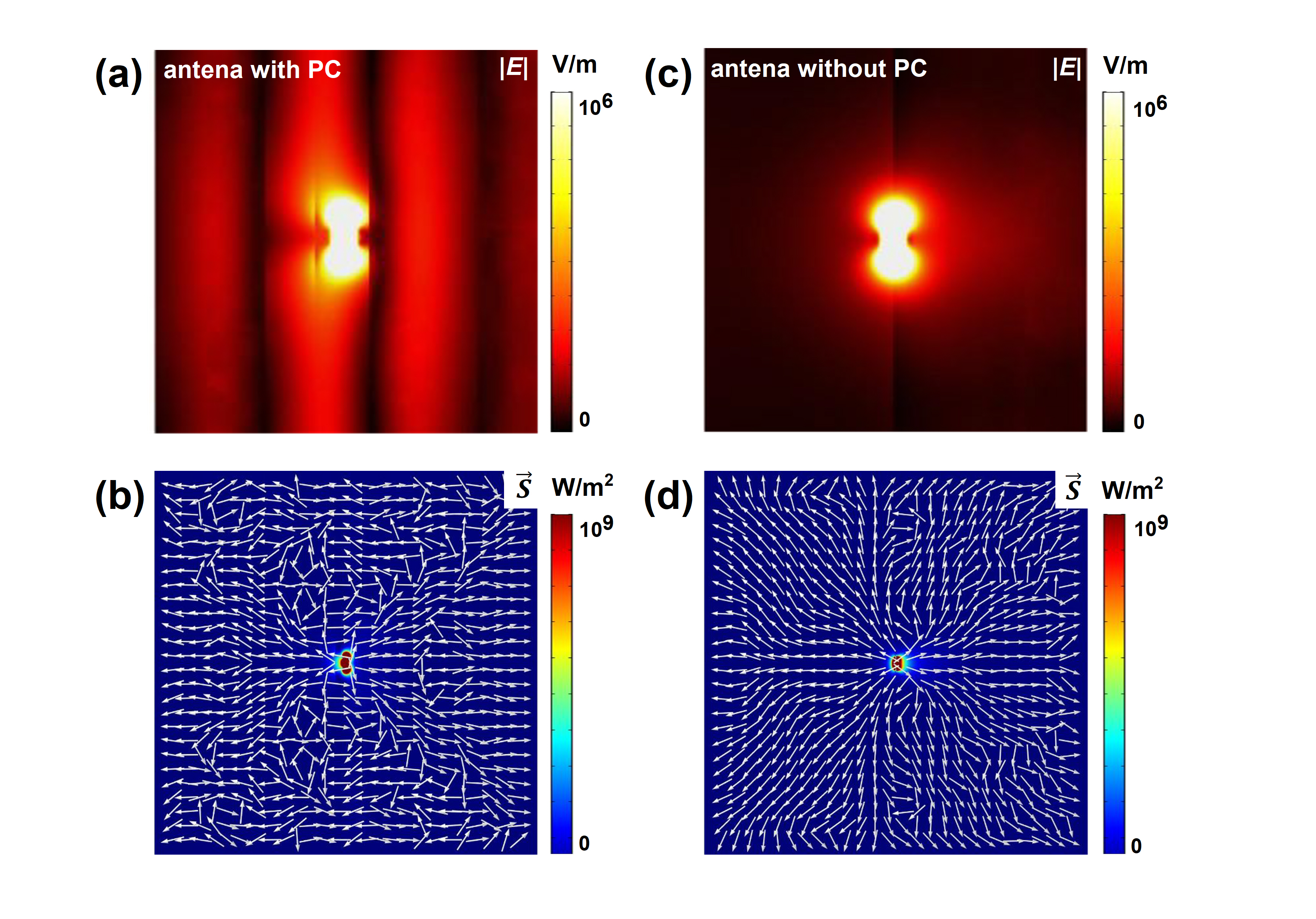}\\
  \caption{The electric field $|E|$ and energy flux $\vec{S}$ distributions (a, b) with and (c, d) without topological PC when the quantum emitter is at the gap center of nanoantenna.
The selected areas are $800\times800$ $nm^2$ for (a, c) and $2.2\times2.2$ $\mu m^2$ for (b, d). 
Comparing with the case without topological PC, the electric field in (a) is pulled away by the edge state and energy flux in (b) is  flowing along $x$ and $-x$ directions rather than scattering to all directions in (d), leading to the  reduction  of  absorption part $\gamma_{ab}$. 
Other parameters are the same as those in Fig. 1(c, d).   
}\label{fig3}
\end{figure}

Next we pay attention to the radios of $\gamma_{ab}/\gamma_{tot}$ and $\gamma_{ed}/\gamma_{tot}$ at the edge state. 
As shown in Tab. 1,  the radio of $ \gamma_{ab}/\gamma_{tot}$ (37.8$\%$ for $r=7$ nm ) in topological structure becomes smaller, and  correspondingly, its edge state part $\gamma_{ed}/\gamma_{tot}$ (62.2$\%$) gets larger,  
comparing with the same size antenna in the dielectric where $ \gamma_{ab}/\gamma_{tot}=45.6\%$  and  $\gamma_{sc}/\gamma_{tot}=54.4\%$ separately. So this is an obvious absorption reduction in Purcell enhancement.

Moreover, absorption reduction with other size of nanoantenna  is also investigated [Tab. 1]. Here the distance $d$ between two nanorods is fixed at $d=10$ nm and resonance matching condition is kept.
First, whether there is a topological structure or not, the ratio of $\gamma_{ab}/\gamma_{tot}$ is smaller when $r$ becomes larger due to less localized field (or larger optical mode volume).
However, for the same $r$, this ratio in topological structure is smaller than that in dielectric medium, namely, it is an obvious absorption reduction in Purcell enhancement.
Especially, with enlarging the nanoantenna, this ratio decreases rapidly [Tab. 1(a)], because the larger deformation of near field caused by the edge state leads to the less energy absorption. 
Therefore, putting the resonant metallic structure into topological structure can effectively reduce the absorption of metallic nanoparticle.
The energy loss is always a drawback in applications of surface plasmon,  using the edge state of tolological structure, it can be effectively overcome.
So this idea can be extended to other topological applications, such as nanolaser and quantum light sources.

\begin{figure}
  \centering
  %Requires \usepackage{graphicx}
  \includegraphics[width=0.5\textwidth]{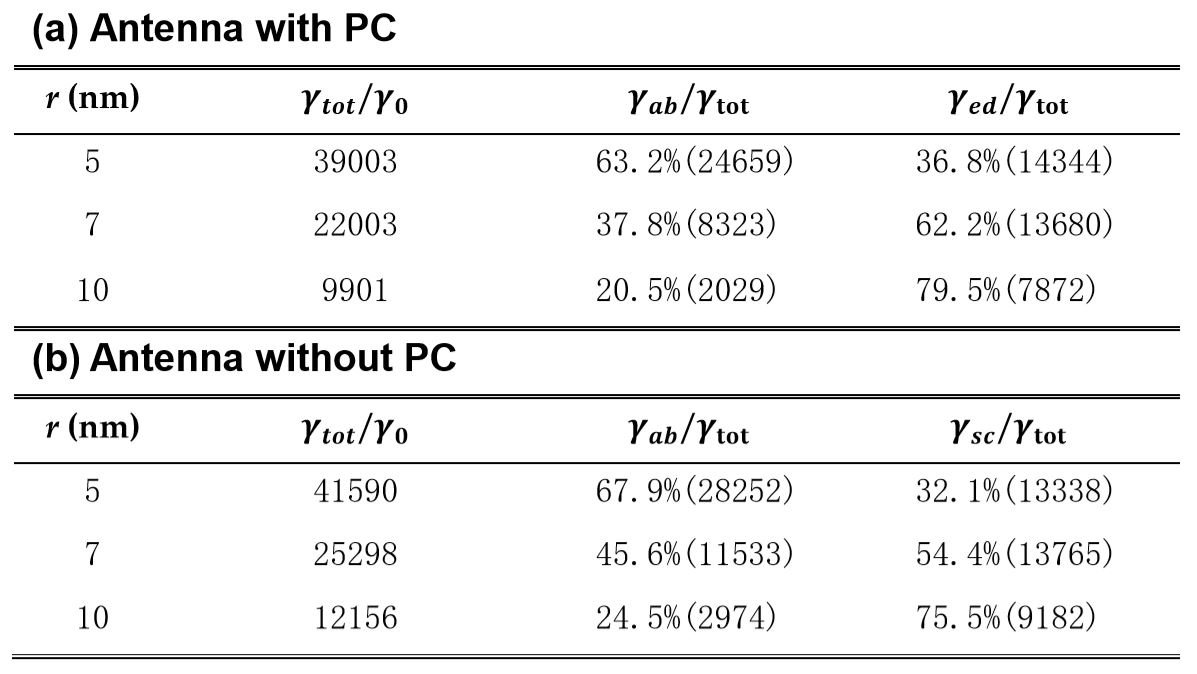}\\
  \caption{  Absorption reduction of Purcell enhancement in topological photonic structure.
  (a) Radios of $\gamma_{ab}/\gamma_{tot}$ and $\gamma_{ed}/\gamma_{tot}$ with topological PC.
  (b) Radios of $\gamma_{ab}/\gamma_{tot}$ and $\gamma_{sc}/\gamma_{tot}$ without PC.
  With the same antenna size, the radio of absorption part  $\gamma_{ab}/\gamma_{tot}$ in PC is less than that in the dielectric. 
  For increasing the size of nanoantenna, this radio is greatly decreased.
  Other parameters are the same as those in Fig. 2(a). 
}\label{fig4}
\end{figure}

Finally, we address fabrication possibility of our scheme. 
%For experimental implementation of our proposal, the fabrication of structure is the key to success. 
Nowadays, nanoantenna \cite{exp-1} and topological PC \cite{exp-2} can be fabricated by state-of-the-art nanotechnology. The single emitter can be realized in many forms, such as classical atoms \cite{exp-3}, Rydberg atoms \cite{exp-4}, and quantum dots \cite{exp-5}. Single emitters embedded in PC waveguide have been realized through scanning tunneling microcopy \cite{exp-6}. The main challenge is to precisely control the nanoantenna at the interface of two PC, which may be solved by atom force microscopy. Thus, it is possible to realize our proposal experimentally in near future.

%%% Section IV
\section{4. Summary}

We have proposed a specific topological photonic structure containing a resonant nanoantenna. 
Under topological protection, extra-large Purcell factors at the edge state as well as the absorption reduction of Purcell enhancement have been achieved.
Owing to without any scattering, these single photons propagating along the edge state can directly be used in on-chip single photon sources and nanolaser.
It  is also promising to extend present 1D topological structure  to  2D topological structures for studying the interaction between photons and quantum emitter.
Nanocavities hiden inside the topological channel pave the way of  utilizing topological photonic structures to realize the behaviors of cavity quantum electrodynamics and quantum information at the micro/nanoscale.

\acknowledgments
We thank  Xiaoyong Hu for helpful discussions. This work is supported by the National Key R$\&$D Program of China under Grant No. 2018YFB1107200, by the National Natural Science Foundation of China under Grants No. 11525414, No. 11974032,  and No. 11734001, and by the Key R$\&$D Program of Guangdong Province under Grant No. 2018B030329001. %Zhichao Li thanks the support from ??.

%\nocite{*}

%\bibliography{EV-whole}

\end{document}